\documentclass[12pt]{article}

\usepackage[margin=1in]{geometry}
\usepackage{amsmath, amssymb, amsthm}
\usepackage{natbib}
\usepackage{hyperref}
\usepackage{enumitem}
\usepackage{booktabs}
\usepackage{xcolor}
\usepackage{setspace}
\usepackage{comment}
\definecolor{cmtcol}{rgb}{0.9, 0.36, 0}

\hypersetup{
  colorlinks=true,
  linkcolor=blue!70!black,
  citecolor=blue!70!black,
  urlcolor=blue!70!black
}

\title{PPI is the Difference Estimator: Recognizing the Survey Sampling Roots of Prediction-Powered Inference}

\author{
  Reagan Mozer\thanks{Department of Mathematical Sciences, Bentley University. Email: \texttt{rmozer@bentley.edu}.}
}

\date{}

\begin{document}
\maketitle

\begin{abstract}
Prediction-powered inference (PPI) is a rapidly growing framework for combining machine learning predictions with a small set of gold-standard labels to conduct valid statistical inference \citep{angelopoulos2023prediction}. In this article, I argue that the core estimators underlying PPI are equivalent to well-established estimators from the survey sampling literature dating back to the 1970s. Specifically, the PPI estimator for a population mean is algebraically equivalent to the difference estimator of \citet{cassel1976some}, and PPI++ corresponds to the generalized regression (GREG) estimator of \citet{sarndal2003model}. 
Recognizing this equivalence, I consider what part of PPI is inherited from a long-standing literature in statistics, what part is genuinely new, and where inferential claims require care. After introducing the two frameworks and establishing their equivalence, I break down where PPI diverges from model-assisted estimation, including differences in the mode of inference, the role of the unlabeled data pool, and the consequences of differential prediction error for subgroup estimands such as the average treatment effect. I then identify what each framework offers the other: PPI researchers can draw on the survey sampling literature's well-developed theory of calibration, optimal allocation, and design-based diagnostics, while survey sampling researchers can benefit from PPI's extensions to non-standard estimands and its accessible software ecosystem. The article closes with a call for integration between these two communities, motivated by the growing use of large language models as measurement instruments in applied research.
\end{abstract}


\section{Introduction}
\label{sec:intro}

Researchers who study text-based outcomes face a familiar bottleneck: every essay, clinical note, or survey response must be read and evaluated by a trained human coder. This problem is not unique to text; for instance, one might want to classify x-ray images, audit financial records, or score video recordings. In each case, measuring the outcome of interest requires expensive, time-consuming expert judgment. Machine learning (ML) models can approximate these measurements cheaply and at scale, but this raises the key question: \textit{is the machine correct?} If the model's predictions contain systematic errors, then using these predictions for downstream statistical analyses can lead to biased results.

This is the motivation behind prediction-powered inference \citep[PPI;][]{angelopoulos2023prediction}, a framework for combining a large set of ML predictions with a small set of gold-standard human labels to conduct valid statistical inference. The key idea is that we can use the labeled data to estimate the prediction error of the ML model, then correct the overall estimate accordingly. The resulting estimator is more efficient than using the labeled data alone and statistically valid regardless of the model's accuracy.

Since its introduction, the PPI framework has rapidly gained traction in the machine learning community. Recent work has included extensions of PPI for settings where predictions are trained on the labeled data \citep[Cross-PPI;][]{zrnic2024cross}, domain-specific extensions for clinical trials \citep{poulet2025ppi} and genomics \citep{miao2024valid}, and for social science applications involving large language models as annotators \citep{egami2024dsl}. 
Despite the growing body of work on PPI, what the literature does not, to my knowledge, fully acknowledge is that the estimators at the core of the framework are just classical tools from an established literature on survey sampling dating back to the 1970s. Indeed, the standard PPI estimator for a population mean is the \emph{difference estimator} of \citet{cassel1976some}, and PPI++ is the \emph{generalized regression (GREG) estimator} of \citet{sarndal2003model}. PPI's central validity guarantee, that inference is valid regardless of the quality of the ML predictions, is the defining property of model-assisted estimation.


I arrived at this observation through my own work on model-assisted estimation for impact analysis in randomized trials with text-based outcomes. In \citet{mozer2025more}, we developed estimators that use machine-predicted text scores to improve precision in randomized experiments, working entirely within the design-based survey sampling framework. 
More recently, \citet{mozer2026stratified} extends this framework to stratified designs and derives the optimal allocation of human coding effort across strata. 
In these papers, the model-assisted estimators we present are structurally equivalent to PPI. 

The purpose of this piece is to formally establish the equivalence between PPI and model-assisted survey estimation and to identify the conditions under which the two coincide versus where they diverge. 
My intention is not to diminish PPI's contributions. 
Rather, I argue that recognizing the survey sampling roots of PPI provides clarity about when and why these estimators work, highlights where assumptions are doing real work, and points to methodological tools the PPI literature can adopt immediately.


The paper proceeds as follows. Section~\ref{sec:estimators} presents the two estimators side by side and establishes their equivalence. Section~\ref{sec:divergences} outlines the key points where the two frameworks diverge. Section~\ref{sec:strengths} discusses the strengths of each framework and what each can offer the other. Section~\ref{sec:crosspollin} closes with a discussion and call to action for better integration and collaboration across the two communities.

\section{The Estimators, Side by Side}
\label{sec:estimators}

\subsection{Setup and notation}

Consider the problem of estimating a population mean $\theta$. A researcher observes a pool of $N$ units for which ML predictions $\hat{Y}_i$ are available, and a labeled subset of $n \ll N$ units for which both the true outcome $Y_i$ and the prediction $\hat{Y}_i$ are observed. To fix ideas, the true outcome might be a human-assigned essay score, and the prediction might be the score assigned by a large language model. Define the prediction error for each labeled unit as $e_i = Y_i - \hat{Y}_i$.

Using only the labeled data, the natural estimator is the sample mean $\hat{\theta}_{\text{lab}} = n^{-1} \sum_{i=1}^n Y_i$, which is unbiased but imprecise when $n$ is small. Using only the predictions, one obtains $\hat{\theta}_{\text{pred}} = N^{-1} \sum_{i=1}^N \hat{Y}_i$, which is precise but potentially biased. The challenge is to combine these two sources of information to get an estimator that is both precise and valid.

\subsection{The PPI estimator}

The PPI estimator \citep{angelopoulos2023prediction} combines these two quantities:
\begin{equation}
\label{eq:ppi}
\hat{\theta}_{\text{PPI}} = \underbrace{\frac{1}{N} \sum_{i=1}^{N} \hat{Y}_i}_{\text{prediction average}} \;+\; \underbrace{\frac{1}{n} \sum_{i=1}^{n} (Y_i - \hat{Y}_i)}_{\text{rectifier}}.
\end{equation}
The second term, called the ``rectifier'', uses the labeled data to correct for systematic prediction error. If the ML model tends to overpredict by, say, half a point on average, the rectifier subtracts that half point. This correction ensures that $\hat{\theta}_{\text{PPI}}$ is unbiased for $\theta$ regardless of how accurate the predictions are.

The PPI++ extension \citep{angelopoulos2023ppi++} introduces a tuning parameter $\lambda$ that controls how much the predictions contribute:
\begin{equation}
\label{eq:ppipp}
\hat{\theta}_{\text{PPI++}} = \frac{1}{n}\sum_{i=1}^n Y_i + \lambda \left(\frac{1}{N}\sum_{i=1}^N \hat{Y}_i - \frac{1}{n}\sum_{i=1}^n \hat{Y}_i\right).
\end{equation}
Setting $\lambda = 1$ recovers the original PPI estimator; setting $\lambda = 0$ ignores the predictions entirely and uses the labeled-only estimator. The optimal $\lambda^*$ is chosen to minimize variance and depends on the predictive accuracy of the ML model. When predictions are good, $\lambda^*$ is close to 1; when they are poor, $\lambda^*$ shrinks toward 0.

\subsection{The model-assisted estimator}

In the survey sampling literature, a researcher has auxiliary information (predictions from a working model) for every unit in a finite population and observes the true ``gold-standard" outcome only for a sampled subset. 
The \textit{difference estimator} \citep{cassel1976some, cassel1977foundations} is defined as
\begin{equation}
\label{eq:diff}
\hat{\theta}_{\text{diff}} = \frac{1}{N}\sum_{i=1}^N \hat{Y}_i + \frac{1}{n}\sum_{i \in s}(Y_i - \hat{Y}_i),
\end{equation}
where $s$ denotes the sample. The \textit{generalized regression} (GREG) estimator \citep{sarndal2003model} generalizes this by weighting the auxiliary information:
\begin{equation}
\label{eq:greg}
\hat{\theta}_{\text{GREG}} = \frac{1}{n}\sum_{i \in s} Y_i + \hat{\beta}\left(\frac{1}{N}\sum_{i=1}^N \hat{Y}_i - \frac{1}{n}\sum_{i \in s} \hat{Y}_i\right),
\end{equation}
where $\hat{\beta}$ is the sample regression coefficient of $Y$ on $\hat{Y}$. The key property of both estimators is \textit{design consistency}: they are unbiased for the population mean regardless of whether the model is correctly specified.

\subsection{The equivalence}

Comparing equations \eqref{eq:ppi} and \eqref{eq:diff} directly: the difference estimator and the PPI estimator are \textit{identical}. Comparing equations \eqref{eq:ppipp} and \eqref{eq:greg}: PPI++ and the GREG estimator are the same formula, with the tuning parameter $\lambda$ playing the role of the regression coefficient $\hat{\beta}$. 

\begin{table}[ht]
\centering
\caption{Concordance between PPI and model-assisted survey estimation: notation, estimator components, and conceptual distinctions.}
\label{tab:notation}
\small
\begin{tabular}{p{3.8cm} p{4.8cm} p{5.2cm}}
\toprule
\textbf{Concept} & \textbf{PPI terminology} & \textbf{Survey sampling terminology} \\
\midrule
Labeled data & Labeled dataset ($n$ units) & Sample $s$ \\[3pt]
Unlabeled data & Unlabeled dataset ($N$ units) & Finite population $\mathcal{U}$ (or first-phase sample in two-phase designs) \\[3pt]
True outcome & Gold-standard label $Y_i$ & Outcome $Y_i$ \\[3pt]
Auxiliary information & Raw inputs (text, images) processed by ML model & Structured covariates $\mathbf{X}_i$ from census or administrative records \\[3pt]
ML/model output & Prediction $\hat{Y}_i$ & Predicted value $m(\mathbf{X}_i; \hat{\boldsymbol{\beta}})$ from working model \\[3pt]
Prediction error & Rectifier component $Y_i - \hat{Y}_i$ & Residual $e_i = Y_i - \hat{Y}_i$ \\[3pt]
Bias correction & Sample mean of rectifier, $n^{-1}\sum_{i \in s}(Y_i - \hat{Y}_i)$ & Sample mean of residuals, $n^{-1}\sum_{i \in s} e_i$ \\[3pt]
Tuning parameter & $\lambda$ (PPI++) & $\hat{\beta}$ (GREG) \\[3pt]
Base estimator ($\lambda=1$) & PPI estimator & Difference estimator \\[3pt]
Tuned estimator & PPI++ & GREG estimator \\[3pt]
Mode of inference & Superpopulation: $(Y_i, \hat{Y}_i) \overset{\text{iid}}{\sim} \mathcal{P}$ & Design-based: randomness from sampling design $p(s)$ \\[3pt]
Inferential target & $\mathbb{E}_\mathcal{P}[Y]$ & $\theta = N^{-1}\sum_{i \in \mathcal{U}} Y_i$ \\[3pt]
Validity guarantee & ``Valid regardless of ML quality'' & Design consistency \\
\bottomrule
\end{tabular}
\end{table}

Table~\ref{tab:notation} maps the terminology and key conceptual distinctions of the two literatures. The point estimators, bias-correction mechanisms, and tuning strategies all correspond to one another. Where the frameworks diverge, in their mode of inference and inferential targets, is the subject of Section~\ref{sec:divergences}.


\section{Where the Two Frameworks Diverge}
\label{sec:divergences}

Although the estimators are algebraically identical, the two frameworks differ in a number of practically meaningful ways in terms of their assumptions, validity, and extensions.

\subsection{Mode of inference and the inferential target}

The most fundamental difference between PPI and model-assisted survey sampling involves the source of randomness. In the design-based framework used in survey sampling, the population $\mathcal{U} = \{1, \ldots, N\}$ is fixed and finite, outcomes $\{Y_1, \ldots, Y_N\}$ are fixed constants, and all randomness comes from the sampling design that determines which units are labeled. The estimand is the finite-population mean $\theta = \frac{1}{N}\sum_{i=1}^N Y_i$. 

In the model-based framework used in PPI, the $N$ units are treated as draws from a superpopulation distribution $\mathcal{P}$, so that $(Y_i, \hat{Y}_i) \overset{\text{iid}}{\sim} \mathcal{P}$, and the $n$ labeled units are a random subsample. Here, the estimand is the superpopulation mean $\theta = \mathbb{E}_\mathcal{P}[Y]$, a parameter of the data-generating distribution.



For the point estimator, this distinction changes nothing: the formula is the same either way. Differences emerge in variance estimation and in what a confidence interval covers.

Under the design-based framework, the variance of the difference estimator under simple random sampling without replacement (SRSWOR) is
\begin{equation}
\label{eq:var_design}
\text{Var}_{\text{des}}(\hat{\theta}_{\text{diff}}) = \left(1 - \frac{n}{N}\right) \frac{S_e^2}{n},
\end{equation}
where $S_e^2 = (N-1)^{-1}\sum_{i=1}^N (e_i - \bar{e})^2$ is the finite-population variance of the prediction residuals $e_i = Y_i - \hat{Y}_i$, and $\bar{e} = N^{-1}\sum_{i=1}^N e_i$ is the population mean residual. The factor $(1 - n/N)$ is the finite-population correction. Notice that the variance depends on the residuals, not on the outcomes themselves: when the predictions $\hat{Y}_i$ are good, the residuals are small, and the variance is low; when the predictions are poor, the residuals are large, and the estimator is no more precise than the simple sample mean of the labeled subset. This is the variance reduction mechanism common to both frameworks.

Under the superpopulation framework used in PPI, $(Y_i, \hat{Y}_i) \overset{\text{iid}}{\sim} \mathcal{P}$, and the variance of the PPI estimator depends on whether the labeled and unlabeled datasets are independent or nested. The default PPI assumption \citep{angelopoulos2023ppi++} treats them as independent draws from $\mathcal{P}$, giving
\begin{equation}
\label{eq:var_ppi}
\text{Var}_{\text{sp}}(\hat{\theta}_{\text{PPI}}) = \frac{\sigma_e^2}{n} + \frac{\sigma_{\hat{Y}}^2}{N},
\end{equation}
where $\sigma_e^2 = \text{Var}_\mathcal{P}(Y - \hat{Y})$ and $\sigma_{\hat{Y}}^2 = \text{Var}_\mathcal{P}(\hat{Y})$. The first term is the cost of estimating the mean residual from a sample of size $n$; the second is the cost of estimating the mean prediction from the unlabeled pool. When the labeled set is instead a subsample of the unlabeled pool (the standard case), the two estimator components are correlated and the variance includes an additional correction term: $\text{Var}(\hat{\theta}) = \sigma_e^2/n + \sigma_{\hat{Y}}^2/N - \sigma_e^2/N + O(n^{-2})$.

In the survey sampling setup, where the predictions are available for the entire finite population, the $\sigma_{\hat{Y}}^2/N$ term vanishes because the mean prediction $\frac{1}{N}\sum_{i=1}^N \hat{Y}_i$ is a known constant. When $N \gg n$, which is typically the case in both literatures, both estimators have a variance of approximately $\sigma_e^2 / n$, with the design-based version $(\hat{\theta}_\text{diff})$ adding a finite-population correction that is itself close to one. \citet{ding2017bridging} formalize this connection, showing that the design-based and superpopulation variance estimators are asymptotically equivalent under simple random sampling with finite fourth moments. 
Thus, in the typical PPI setting (e.g., $n$ in the hundreds, $N$ in the thousands or tens of thousands), the practical differences between the two formulas are negligible. 

Where the two frameworks diverge more meaningfully is in what the inferential target means, particularly for causal inference. 
Consider, for example, a randomized experiment with a finite population of $N$ units. Under the design-based framework, the estimand is the \textit{sample average treatment effect} (SATE), $\tau_{\text{SATE}} = N^{-1}\sum_{i=1}^N [Y_i(1) - Y_i(0)]$, which describes the average effect for exactly these $N$ individuals. Under the superpopulation framework, the estimand is the \textit{population average treatment effect} (PATE), $\tau_{\text{PATE}} = \mathbb{E}_\mathcal{P}[Y(1) - Y(0)]$, which describes the expected effect for a hypothetical population from which the $N$ experimental units were drawn. The SATE is a fixed constant; the PATE is a parameter of a probability model. Both are sensible targets, but they answer different questions, and confidence intervals for one do not necessarily have correct coverage for the other.

In small studies or experiments with heterogeneous treatment effects, the difference can be practically meaningful.
 The PATE variance includes terms for between-subject variability in treatment effects that are absent from the SATE variance. For a researcher using PPI to estimate a treatment effect in a specific randomized trial, the design-based (SATE) framing may be more natural. For a researcher using PPI to draw conclusions about a broader population from which the study sample was drawn, the superpopulation (PATE) framing may be more appropriate. Neither is wrong, but the choice should be intentional rather than implicit.

\subsection{The role of the working model}
\label{subsec:working_model}

Both frameworks require predicted values $\hat{Y}_i$ for every unit in the population or unlabeled pool, but they arrive at these predictions in different ways.

In model-assisted estimation under the survey sampling framework, the predictions come from an explicit \textit{working model}\footnote{Here, the ``working'' term reflects that the model need not be correctly specified: validity rests on the sampling design, not the model.} that relates the outcome $Y$ to a set of auxiliary variables $\mathbf{X}$ observed for every unit in the population. These auxiliary variables are typically census-level covariates (e.g., demographic characteristics, geographic variables, or historical data collected from administrative records). The working model has a specified functional form, $\hat{Y}_i = m(\mathbf{X}_i; \hat{\boldsymbol{\beta}})$, where $m$ might be a linear regression, a spline, or a random forest, and $\hat{\boldsymbol{\beta}}$ are parameters estimated from the sample \citep{sarndal2003model, breidt2017model}. 


In PPI, the predictions $\hat{Y}_i$ are typically outputs of a ML model or large language model trained to approximate the outcome of interest. There is no requirement to decompose the prediction mechanism into covariates and a functional form. Rather, the prediction model is treated as a black box: it takes in raw data (e.g., the full text of an essay, an image, a clinical note) and generates a prediction. The PPI framework takes those predictions as given and asks only whether the labeled data are sufficient to correct whatever errors the predictions contain.

Both frameworks ultimately require population-level auxiliary data, but where the two setups genuinely differ is in the \textit{nature} of this auxiliary information and how they handle data-dependent predictions. In survey sampling, this comes in the form of structured covariates collected from administrative records or census data. In PPI applications with unstructured data, the auxiliary information is the raw inputs themselves (e.g., raw texts or images). The ML model converts these into scalar predictions, providing the kind of auxiliary variable that the model-assisted framework needs. 

The difference in how predictions are obtained has several consequences. First, in the survey sampling setup, the working model is typically fitted using the same labeled sample that will be used for estimation. This creates a dependence between $\hat{Y}_i$ and the sample $s$ that must be handled carefully in variance estimation. The classical model-assisted theory says that even when $\hat{\boldsymbol{\beta}}$ is sample-dependent, the resulting estimator is still asymptotically design-consistent, a result that holds because the model is ``assisted'' rather than ``depended on'' \citep{sarndal2003model}. In PPI, the default assumption is that the ML model is trained on an external dataset, so the predictions are fixed quantities that do not depend on the labeled sample. When this assumption is violated, as when the model is trained or fine tuned using the subset of labeled data, Cross-PPI \citep{zrnic2024cross} provides a straightforward sample-splitting correction. Both frameworks thus handle data-dependent predictions, but through different mechanisms.

Second, the black-box nature of ML predictions in the PPI setting makes it harder to diagnose \textit{where} and \textit{why} differential  prediction errors arise. In the survey sampling framework, because the working model has an explicit functional form, a researcher can evaluate the model residuals, check for heteroscedasticity across subgroups, or examine whether particular covariates are driving poor predictions. In the PPI setting, diagnosing differential prediction error requires comparing residuals within the labeled sample across existing groups without knowledge of the underlying model structure. See Section~\ref{subsec:diff_pred_error} for further discussion of this issue.

\subsection{The role of the unlabeled data pool}

In survey sampling, the predictions $\hat{Y}_i$ are available for every unit in the finite population, treated as census-level auxiliary data. The labeled subset is sampled from this population, so the remaining unlabeled units are simply the complement of the sample. Thus, the prediction-only average $\frac{1}{N}\sum_{i=1}^N \hat{Y}_i$ is a known population quantity.

In PPI, the unlabeled dataset is treated as a sample from an infinite superpopulation. In practice, the unlabeled pool is often a convenience sample or a moving stream updated over time. 
The $N$ unlabeled units and $n$ labeled units may overlap (with the labeled units being a subsample of the unlabeled pool) or may be drawn independently. The flexibility to draw on labeled data from a separate pool is potentially very useful, but it creates an important question: \textit{does inference on the unlabeled pool target the same thing as the labeled subset?} 

While PPI allows for the labeled sample to be disjoint from the unlabeled pool (with both viewed as draws from the same population distribution), the standard PPI variance formula does not account for the additional uncertainty from the unlabeled pool being a sample rather than a census. The survey sampling framework handles this using variance estimators that incorporate both phases of sampling. In survey sampling, this is known as a two-phase design, where a large first-phase sample collects inexpensive auxiliary data, and a smaller second-phase subsample is selected to measure the expensive outcome \citep{sarndal2003model}. 

\subsection{Differential prediction error and treatment effects}
\label{subsec:diff_pred_error}

For a population mean under simple random sampling, both frameworks lead to identical estimators with the same guarantees. However, in the PPI framework, problems can arise when the residual correction is not exchangeable across subgroups. A practically important case is treatment-effect estimation with machine-scored outcomes.

Suppose a researcher uses an ML model to score a text outcome (e.g., essay quality) and estimates a treatment effect $\tau = \bar{Y}_1 - \bar{Y}_0$ by applying PPI with a single pooled rectifier computed from all labeled data. Let $\delta_z = \mathbb{E}[Y - \hat{Y} \mid Z = z]$ denote the arm-specific mean prediction error. Because the pooled rectifier estimates a weighted average of $\delta_0$ and $\delta_1$, it cancels in the difference between arms, giving
\begin{equation}
\label{eq:bias_tau}
\text{Bias}(\hat{\tau}) = \delta_0 - \delta_1
\end{equation}
(see Appendix~\ref{app:bias_derivation} for the full derivation). Importantly, this bias depends only on the differential prediction error across treatment arms and is completely unaffected by the size of the labeled sample (i.e., collecting more labeled data does not reduce the magnitude of bias).

This issue is particularly important when LLMs are used to obtain predictions, as LLMs are often especially vulnerable to differential prediction error. For instance, interventions that change the content of text responses (e.g., a writing curriculum that produces more structured, evidence-based essays) may alter the distribution of text features in a way that affects accuracy, leading the LLM to score treated essays more reliably than control essays, or vice versa. LLM outputs are also sensitive to prompt wording, so texts near the decision boundary of the scoring rubric may exhibit higher variability. Thus, an intervention that shifts texts toward or away from that boundary may lead to arm-specific noise. In addition, LLM providers often update their models without notice. If labeled data are scored by one version of an LLM during and unlabeled data are scored by a later version, the prediction errors are not exchangeable across the two datasets.

The fix is to apply the PPI rectifier separately within each treatment arm, so that arm-specific prediction biases are corrected by arm-specific rectifiers. This is the classical non-differential measurement error principle \citep{carroll2006measurement} in another form.  Within the survey sampling framework, estimating separately within subpopulations is a standard practice, and this stratified approach would be a natural default. In the PPI literature, where the rectifier is presented as a single correction applied to the full sample, this recommendation is largely absent.


\section{What Each Framework Offers the Other}
\label{sec:strengths}

\subsection{What the survey sampling framework offers PPI}
The survey sampling community has spent decades developing tools that address problems the PPI community is only just beginning to encounter. Below, I highlight four areas where the existing literature is directly applicable to PPI.


\paragraph{Calibration and post-stratification.} 

The calibration framework of \citet{deville1992calibration} adjusts estimation weights so that sample-level covariate distributions match known population distributions. In PPI terms, this means ensuring that the distribution of ML predictions in the labeled subsample matches their distribution in the full pool. Post-stratification, forming strata based on predicted values and computing weighted within-stratum estimates, can improve precision beyond what unstratified PPI achieves. \citet{miratrix2013adjusting} provide formal results on efficiency gains from post-stratification in randomized experiments that are directly relevant to PPI users estimating treatment effects and \citet{mozer2026stratified} discusses extensions of these methods specifically designed for text-as-data applications.

\paragraph{Optimal allocation of labeling effort.} 

One of the most immediate practical questions a PPI user faces is: given a fixed budget for human labeling, which units should be sent to coders? Survey sampling has a well-developed theory for exactly this problem. The classical Neyman allocation \citep{Neyman34} approach says to label more units in regions where prediction errors are most variable. \citet{mozer2026stratified} show that strategic allocation of human labeling effort across strata defined by ML or LLM-generated predictions can provide meaningful efficiency gains beyond what model-assisted estimation alone provides. Related work on optimal allocation in factorial experiments \citep{ravichandran2024optimal} provides additional tools for settings where labeling budgets must be divided across treatment conditions. The emerging ``active PPI'' literature is moving toward this direction, but the survey sampling literature provides established optimality results that can serve as a starting point for PPI extensions.


\paragraph{Design-based diagnostics.} 
The survey sampling framework provides a clear workflow for diagnostic checks that are, to my knowledge, absent from the PPI literature. Researchers implementing PPI should ask: is the distribution of prediction errors in the labeled subsample representative of prediction errors in the full pool? Do prediction errors vary systematically across subgroups of interest, such as treatment arms? What is the design effect \citep{kish1965survey}, the ratio of the PPI estimator's variance to the labeled-only estimator's variance? 

\citet{sarndal2003model} develop residual-based diagnostics for assessing whether the working model's predictions are well-behaved across subpopulations, and \citet{valliant2000finite} frame the entire estimation problem through the lens of prediction residuals. In causal inference applications, a useful check is to compare the distribution of residuals between treated and control units in the labeled sample; large differences are an early warning of differential prediction error. These diagnostics are standard in survey sampling methodology but seemingly absent from PPI software.

\paragraph{Sensitivity analysis for assumption violations.} 

The survey sampling and measurement error literatures \citep{carroll2006measurement} provide established frameworks for assessing sensitivity to the assumptions underlying PPI. The most critical assumption is that prediction errors are exchangeable between the labeled and unlabeled data. In causal inference applications, an additional concern is that prediction errors do not vary systematically across treatment conditions. Rosenbaum's sensitivity analysis framework for hidden bias in observational studies \citep{rosenbaum2002observational} can be adapted to this setting to determine the degree of differential prediction error that would be needed to overturn a finding.
Similarly, the SIMEX approach of \citet{cook1994simulation} provides a simulation-based method for tracing how estimates change as prediction errors become more unstable. Following \citet{lipsitch2010negative}, one could also apply the PPI treatment effect estimator to outcomes where no treatment effect is expected. Here, a nonzero result would suggest that prediction errors are not exchangeable across treatment arms.

\subsection{What PPI offers the survey sampling community}
The survey sampling community also has much to gain from engaging with PPI and its extensions.

\paragraph{Flexible estimands.}
The PPI framework is designed to handle a broad class of estimands beyond population means. \citet{angelopoulos2023prediction} provide a general recipe: any parameter $\theta^*$ that can be expressed as the minimizer of a convex loss function $L(\theta) = \mathbb{E}[\ell_\theta(X, Y)]$ can be estimated via PPI by replacing the population loss with a ``rectified'' version that combines predictions on unlabeled data with a bias correction from labeled data. This covers linear and logistic regression coefficients, quantiles, and general M-estimators. PPI++ \citep{angelopoulos2023ppi++} extends this with computationally efficient algorithms for generalized linear models, and Cross-PPI \citep{zrnic2024cross} provides the same recipe to handle data-dependent predictions.


\paragraph{Tuning parameter selection.}
A key practical question for both frameworks is how much weight to place on the ML predictions relative to the labeled data alone. For the GREG estimator, the weight placed on the auxiliary information is the regression coefficient $\hat{\beta}$, estimated as part of the model fitting procedure but rarely discussed as a tuning choice. PPI++ reframes this as an explicit tuning parameter $\lambda \in \mathbb{R}$ with a closed-form optimum $\lambda^* = \text{Cov}(Y, \hat{Y}) / \text{Var}(\hat{Y})$ (for mean estimation), which gives researchers a dial for controlling the contribution of ML predictions as well as an interpretable measure of whether the predictions are actually helping.

\paragraph{Cross-fitting for sample-dependent predictions.} 
Cross-PPI \citep{zrnic2024cross} handles the setting where the ML model is trained on the labeled data, using sample-splitting to avoid overfitting bias. This is common in practice: a researcher trains a classifier on hand-coded essays, then uses it to predict scores for the rest. The survey sampling literature addresses related issues through the jackknife and bootstrap \citep{rao1988resampling, sarndal2003model}, but Cross-PPI packages the solution in a natural and straightforward way.

\paragraph{Accessible software.} The \texttt{ppi\_py} Python package and the \texttt{dsl} R package from \citet{egami2024dsl} make PPI estimators accessible and easy to implement in standard workflows.  Software for implementing survey sampling methods, by contrast, is less unified, with different implementations spread across various specialized packages in R (e.g., \texttt{survey}; \citealt{lumley2004analysis}). A PPI-style interface built on top of the \texttt{survey} R package would make model-assisted estimation far more accessible to applied researchers in the ML community.


\section{Discussion}
\label{sec:crosspollin}

Within both the machine learning and survey sampling communities, researchers have arrived at the same solution for the problem of combining cheap, imperfect proxy outcomes with expensive, gold-standard labels. For mean estimation, PPI and model-assisted survey estimation are two presentations of the same estimator. The fact that both communities arrived at the same answer separately and continue to develop in relative isolation is, I believe, a real cost. 

For PPI researchers, the survey sampling literature offers immediate practical benefits: exact variance formulas under various sampling designs, optimality results for allocation, well-studied calibration procedures, and diagnostic tools for assessing when predictions are helping and when they are not. Researchers adopting PPI should understand that they are working with a class of estimators whose properties have been studied for fifty years, and leveraging this knowledge can save considerable effort moving forward.

For survey sampling researchers, PPI offers a modern take on model-assisted estimation that can appeal to a broader audience. The emphasis on accessible software, clear-cut recipes for inference with flexible estimands, and practical tuning procedures shows how classical ideas can be repackaged for modern analysis workflows. PPI's extensions to non-smooth estimands and the Cross-PPI framework for data-dependent predictions address settings that are increasingly relevant in applied survey work but have received little attention to date.

Below, I provide four concrete suggestions for researchers in both fields to navigate the path forward:
\begin{enumerate}
\item Treat PPI estimators as members of a broader model-assisted family, and report design assumptions as explicitly as model assumptions.
\item Bring survey diagnostics, calibration, and post-stratification tools into standard PPI implementations.
\item Import PPI developments for flexible estimands into survey-sampling.
\item Evaluate both classes of estimators across common benchmarks where the inferential target, labeling mechanism, and unlabeled-pool construction are explicit.
\end{enumerate}


As large language models become the dominant tool for annotating unstructured data, the need for principled frameworks that leverage ML predictions while maintaining valid inference will only grow.
Researchers from both communities stand to gain from treating this as one common methodological goal.

In closing, my call to researchers in the PPI and survey sampling domains is this: the estimators are the same. The conversation should be too.


\bibliographystyle{apalike}
\bibliography{refs}

\appendix
\section{Derivation of Treatment Effect Bias Under Differential Prediction Error}
\label{app:bias_derivation}

Consider a randomized experiment with binary treatment $Z \in \{0,1\}$. Let $N_z$ denote the number of units assigned to arm $z$, and suppose that a labeled subsample of size $n_z$ is drawn uniformly at random within each arm. Write $f_z = n_z / N_z$ for the labeling fraction in arm $z$, and let $S_z$ and $U_z = \{1,\dots,N_z\} \setminus S_z$ denote the labeled and unlabeled index sets, respectively. For every unit $i$ in arm $z$, let $Y_i$ be the true outcome and $\hat{Y}_i$ a prediction from an externally trained model. Define the arm-specific mean prediction error as
\[
\delta_z \;=\; \mathbb{E}[Y_i - \hat{Y}_i \mid Z_i = z],
\]
so that $\mathbb{E}[\hat{Y}_i \mid Z_i = z] = \mu_z - \delta_z$, where $\mu_z = \mathbb{E}[Y_i \mid Z_i = z]$ is the true arm mean. The true treatment effect is $\tau = \mu_1 - \mu_0$.

\subsection*{PPI with a pooled rectifier}

Suppose the researcher applies the PPI estimator to each arm using a single rectifier computed from all labeled data pooled across arms. Let $S = S_0 \cup S_1$ denote the full labeled set, with $n = n_0 + n_1$ total labeled units. The arm-specific PPI estimate with pooled rectifier is
\begin{equation}
\hat{\theta}_z^{\text{pool}} \;=\; \bar{\hat{Y}}_{N_z} + \bigl(\bar{Y}_{S} - \bar{\hat{Y}}_{S}\bigr),
\label{eq:pooled_ppi}
\end{equation}
where $\bar{\hat{Y}}_{N_z} = N_z^{-1}\sum_{i=1}^{N_z}\hat{Y}_i$ is the full-arm prediction mean, and $\bar{Y}_S$ and $\bar{\hat{Y}}_S$ are the labeled-set means of true outcomes and predictions, respectively, pooled across both arms. The treatment effect estimate is $\hat{\tau}^{\text{pool}} = \hat{\theta}_1^{\text{pool}} - \hat{\theta}_0^{\text{pool}}$.

Taking expectations, $\mathbb{E}[\bar{\hat{Y}}_{N_z}] = \mu_z - \delta_z$. The pooled rectifier has expectation
\[
\mathbb{E}[\bar{Y}_S - \bar{\hat{Y}}_S] = \frac{n_0}{n}\,\delta_0 + \frac{n_1}{n}\,\delta_1 \;\equiv\; \bar{\delta},
\]
a weighted average of the arm-specific prediction errors. Therefore,
\[
\mathbb{E}[\hat{\theta}_z^{\text{pool}}] = (\mu_z - \delta_z) + \bar{\delta} = \mu_z - (\delta_z - \bar{\delta}).
\]
The pooled rectifier corrects toward $\bar{\delta}$, but the arm-specific error $\delta_z$ may differ from it. Taking the difference,
\begin{align}
\mathbb{E}[\hat{\tau}^{\text{pool}}]
&= \bigl[\mu_1 - (\delta_1 - \bar{\delta})\bigr] - \bigl[\mu_0 - (\delta_0 - \bar{\delta})\bigr] \notag \\
&= (\mu_1 - \mu_0) - \delta_1 + \delta_0 \notag \\
&= \tau + \delta_0 - \delta_1.
\label{eq:pooled_bias}
\end{align}
The weighted average $\bar{\delta}$ cancels in the difference, leaving
\[
\text{Bias}(\hat{\tau}^{\text{pool}}) = \delta_0 - \delta_1.
\]
This bias depends only on the differential prediction error across arms. It is unaffected by the labeling rate and cannot be reduced by collecting more labeled data.

\subsection*{Naive prediction imputation (no rectifier)}

A researcher who forgoes the rectifier entirely and simply substitutes predictions for unlabeled outcomes faces an even larger bias. Define the naive arm-specific estimator as
\begin{equation}
\hat{\theta}_z^{\text{naive}} \;=\; f_z \, \bar{Y}_{S_z} \;+\; (1 - f_z)\,\bar{\hat{Y}}_{U_z},
\label{eq:naive_estimator}
\end{equation}
where $\bar{Y}_{S_z} = n_z^{-1}\sum_{i \in S_z} Y_i$ and $\bar{\hat{Y}}_{U_z} = (N_z - n_z)^{-1}\sum_{i \in U_z} \hat{Y}_i$. Taking expectations,
\begin{align}
\mathbb{E}[\hat{\theta}_z^{\text{naive}}]
&= f_z\,\mu_z + (1 - f_z)\,(\mu_z - \delta_z) \notag \\
&= \mu_z - (1 - f_z)\,\delta_z.
\label{eq:naive_arm_expectation}
\end{align}
The treatment effect bias is then
\begin{align}
\text{Bias}(\hat{\tau}^{\text{naive}})
&= \mathbb{E}[\hat{\theta}_1^{\text{naive}} - \hat{\theta}_0^{\text{naive}}] - \tau \notag \\
&= (1 - f_0)\,\delta_0 - (1 - f_1)\,\delta_1.
\label{eq:naive_te_bias}
\end{align}
Under balanced labeling ($f_0 = f_1 = f$), this simplifies to $(1 - f)(\delta_0 - \delta_1)$, which converges to $\delta_0 - \delta_1$ as $f \to 0$ and vanishes as $f \to 1$. The naive estimator thus combines the differential prediction error with an additional penalty that grows as the labeling rate decreases.

\subsection*{Eliminating the bias via arm-specific rectifiers}

The PPI framework, equivalently the difference estimator of \citet{cassel1976some}, eliminates both sources of bias by applying a separate rectifier within each treatment arm:
\[
\hat{\theta}_z^{\text{PPI}} = \bar{\hat{Y}}_{N_z} + \bigl(\bar{Y}_{S_z} - \bar{\hat{Y}}_{S_z}\bigr),
\]
where $\bar{\hat{Y}}_{S_z} = n_z^{-1}\sum_{i \in S_z}\hat{Y}_i$ is the labeled-set prediction mean \textit{within arm $z$}. The rectifier $\bar{Y}_{S_z} - \bar{\hat{Y}}_{S_z}$ is an unbiased estimate of $\delta_z$, which exactly cancels the arm-specific prediction error. The resulting treatment effect estimator $\hat{\tau}^{\text{PPI}} = \hat{\theta}_1^{\text{PPI}} - \hat{\theta}_0^{\text{PPI}}$ is unbiased for $\tau$ regardless of the prediction quality or the labeling fractions.

\end{document}